\pdfoutput=1
\documentclass[12pt]{article}
\usepackage{amsmath,amssymb,amsfonts}
\usepackage[multiple,stable]{footmisc}
\usepackage[amsmath,amsthm,thmmarks]{ntheorem}
\allowdisplaybreaks[4]
\usepackage[noconfig]{refstyle}
\usepackage[dvipsnames]{xcolor}
\usepackage[hyperfootnotes=false, linktocpage=true, colorlinks, citecolor=blue, linkcolor=blue, urlcolor=Maroon]{hyperref}
\usepackage{array,tabularx,booktabs,geometry,subfigure,graphicx,longtable}
\usepackage[bf]{caption}
\usepackage{tikz}
\usetikzlibrary{shapes,arrows}
\tikzstyle{block} = [rectangle, draw, text width=7em, text centered, rounded corners, minimum height=3em]
\usepackage{graphicx,color}
\usepackage{cite}

\let\eqref=\relax
\newref{eq}{name={eq.~},Name={Eq.~},names={eqs.~},Names={Eqs.~},rngtxt={-},refcmd=(\ref{#1})}
\newref{f}{name={footnote~},Name={Footnote~},names={footnotes~},Names={Footnotes~}}
\newref{app}{name={appendix~},Name={Appendix~},names={appendixes~},Names={Appendixes~}}
\newref{tab}{name={table~},Name={Table~},names={tables~},Names={Tables~}}
\newref{ch}{name={chapter~},Name={Chapter~},names={chapters~},Names={Chapters~}}
\newref{sec}{name={section~},Name={Section~},names={sections~},Names={Sections~}}
\newref{fig}{name={figure~},Name={Figure~},names={figures~},Names={Figures~}}
\numberwithin{equation}{section}
\newcommand{\eref}[1]{(\ref{#1})}

\newcommand{\eeq}{\end{equation}}
\newcommand{\beq}{\begin{equation}}
\newcommand{\ba}{\begin{array}}
\newcommand{\ea}{\end{array}}

\newcommand{\cM}{{\cal M}}

\newcommand{\cO}{{\cal O}}

\newcommand{\IP}{\mathbb P}
\newcommand{\IC}{\mathbb C}

\newcommand{\be}{\begin{equation}}
\newcommand{\ee}{\end{equation}}
\newcommand{\bea}{\begin{equation}\begin{aligned}}	
\newcommand{\eea}{\end{aligned}\end{equation}}		

\newcommand{\iddots}{\mathinner{\mkern2mu\raise1pt\hbox{.}\mkern2mu \raise4pt\hbox{.}\mkern2mu\raise7pt\hbox{.}\mkern1mu}}

\providecommand{\id}{\leavevmode\hbox{\small$\mathrm{1}$\kern-3.8pt\normalsize$\mathrm{1}$}}
\def\fnote#1#2{\begingroup\def\thefootnote{#1}\footnote{#2}
     \addtocounter{footnote}{-1}\endgroup}

\begin{document}

\vspace{1cm}

\title{
       \vskip 40pt
       {\huge \bf On Instanton Superpotentials, Calabi-Yau Geometry, and Fibrations}}

\vspace{2cm}

\author{Lara B. Anderson${}^{1}$, Fabio Apruzzi${}^{2,3,4}$, Xin Gao${}^{1}$, James Gray${}^{1}$ and Seung-Joo Lee${}^{1}$}
\date{}
\maketitle
\begin{center} {\small ${}^1${\it Physics Department, Robeson Hall, Virginia Tech, Blacksburg, VA 24061, USA}}\\
{\small ${}^{2}${\it University of North Carolina, Department of Physics and Astronomy, Phillips Hall, CB $\#$3255, 120 E. Cameron Ave., Chapel Hill, NC 27599-3255, USA}}\\
{\small ${}^{3}${\it City University of New York, The Graduate Center, 365 Fifth Avenue, New York, NY 10016, USA}}\\
{\small ${}^{4}${\it Columbia University, Department of Physics, Pupin Hall, 550 West 120th St., New York, NY, 10027, USA}}\\
\fnote{}{lara.anderson@vt.edu, fabio.apruzzi@unc.edu, xingao@vt.edu, jamesgray@vt.edu, seungsm@vt.edu}
\end{center}

\begin{abstract}
\noindent
In this paper we explore contributions to non-perturbative superpotentials arising from instantons wrapping effective divisors in smooth Calabi-Yau four-folds. We concentrate on the case of manifolds constructed as complete intersections in products of projective spaces (CICYs) or generalizations thereof (gCICYs). We systematically investigate the structure of the cone of effective (algebraic) divisors in the four-fold geometries and employ the same tools recently developed in \cite{Anderson:2015iia} to construct more general instanton geometries than have previously been considered in the literature. We provide examples of instanton configurations on Calabi-Yau manifolds that are elliptically and $K3$-fibered and explore their consequences in the context of string dualities. The examples discussed include manifolds containing infinite families of divisors with arithmetic genus, $\chi(D, \cO_D)=1$ and superpotentials exhibiting modular symmetry. \end{abstract}

\thispagestyle{empty}
\setcounter{page}{0}
\newpage

\tableofcontents

\section{Introduction} \label{intro}
In the search for stabilized vacua and realistic models of string phenomenology, non-perturbative effects are an essential ingredient. In particular, instanton contributions have played a significant role in most attempts at model building and moduli stabilization in realistic $4$-dimensional, $\mathcal N=1$ compactifications. Examples of detailed calculations of instanton effects can be found in \cite{Beasley:2003fx,Braun:2007xh,Diaconescu:1999it,Braun:2007tp,Braun:2007vy,Curio:2010hd,Curio:2009wn,Curio:2008cm} (heterotic string theory), \cite{Friedman:1997yq,Donagi:1996yf,Braun:2000hh,Curio:1997rn,Cvetic:2012ts} (F-theory) and \cite{Font:1990gx,Grimm:2007xm,Ferrara:1989bc,Beasley:2005iu} (Type II).

In order to fully employ these non-perturbative effects it is clear that systematic/algorithmic control of the underlying geometry is essential. In this paper, we will develop a partial toolkit for such an approach and look at one particular window into these non-perturbative effects in a network of dual theories. We will explore the geometry associated with M-theory instantons on a class of smooth Calabi-Yau (CY) four-folds \cite{Brunner:1996bu,Gray:2013mja,Gray:2014kda,Gray:2014fla}, $Y_4$, constructed as complete intersections in simple projective ambient spaces \cite{Hubsch:1986ny,Candelas:1987kf,Green:1986ck,Candelas:1987du}, as well as recent generalizations \cite{Anderson:2015iia} of this construction. Many of our results readily extend to other constructions of CY four-folds such as \cite{Morrison:2014lca,Anderson:2014gla,Klemm:1996ts,Kreuzer:1997zg,Lynker:1998pb,Halverson:2015jua}. It is well established that very simple observations regarding these non-perturbative effects -- including the structure of the complex divisors, $D \subset Y_4$, wrapped by $M5$-branes -- can have a wide range of consequences. This is true not only for the effective 3-dimensional description of M-theory compactified on $Y_4$, but also for many other theories related by string dualities (i.e. Heterotic, Type IIB and F-theory) \cite{Friedman:1997yq}. 

It should be noted that here we will consider only the most universal sector of instanton contributions and the geometry of smooth Calabi-Yau four-folds. Within the context of realistic string compactifications and dual theories (especially heterotic, F-theory and Type IIB) this is only a first step. In the language of F-theory for example, the analysis presented here pertains only to ED3-ED3 instanton zero modes and omits the important consideration of ED3-7 instanton zero modes which also play a crucial role. There is a rich literature on this subject including investigations of the cohomology on divisors \cite{Cvetic:2010rq,Cvetic:2010ky,Cvetic:2011gp,Blumenhagen:2010ja}, lifting of zero modes, and the role of fluxes and $U(1)$ symmetries \cite{Grimm:2011dj,Bianchi:2012pn,Martucci:2015oaa,Martucci:2015dxa} in these questions. We view the geometric tools for smooth CY four-folds explored here as essential but only preliminary steps towards a comprehensive study of instantons in realistic heterotic/F-theory/Type IIB vacua.

In particular, one of our primary goals in this work is to explore fibration structures and effective divisors on $Y_4$ that are not manifestly ``inherited" from the ambient space. That is, for CY four-folds described in some simple ambient space, $Y_4 \subset {\cal A}$, we are interested in effective divisors $D \subset Y_4$ that are not the restriction, ${\cal D}|_{Y_4}$, of some effective divisor ${\cal D} \subset {\cal A}$. It is this latter type of divisor that has been most frequently used to explore instanton solutions in the literature. As an illustration of why it is necessary to study more general solutions, it should be noted that reference \cite{Witten:1996bn} demonstrated that, for some classes of ordinary complete intersection manifolds in products of projective spaces, no inherited divisors can satisfy the necessary conditions to contribute non-trivially to the superpotential. 

The goals of the present work include:
\begin{itemize}
\item A complete, systematic study of the structure of effective cones of divisors on $Y_4$ and an investigation of general divisors with arithmetic genus $1$ in smooth CY four-folds. 
\item An exploration of the structure of instantons on four-folds constructed as ``Generalized Complete Intersections" (i.e. ``gCICYs") \cite{Anderson:2015iia} in compactifications of M-theory. 
\item An investigation of the consequences of instantons wrapping ``non-inherited" divisors in string dualities -- including infinite families of instanton solutions exhibiting modular symmetry.
\end{itemize}

With these goals in mind, we turn first to the essential mathematical structure that we wish to explore in this work: effective divisors on $Y_4$ that are \emph{not} inherited from ${\cal A}$.

\subsection{Exploring the full cone of effective divisors}\label{first_tools}
In this section we explore the simple geometric fact that the cone of effective divisors (i.e. the co-dimension $1$ algebraic sub-varieties of $Y_4$) can be significantly larger than that ``inherited" from the ambient space. As a straightforward example, consider the CY four-fold, described via a degree $(2,5)$ hypersurface in the product of complex projective spaces $\mathbb{P}^1 \times \mathbb{P}^4$:
\begin{eqnarray} \label{first_eg}
Y_4 = \left[ \begin{array}{c||c}
 \mathbb{P}^1 & 2 \\ 
\mathbb{P}^4 & 5 \\ 
\end{array}\right]\;.
\end{eqnarray}

Since this manifold is defined via an ample hypersurface, its $H^{1,1}$ cohomology group descends simply from the ambient product of projective spaces. Here $h^{1,1}(Y_4)=2$ and a basis of the Picard group is given by $H_1$ and $H_2$, the restrictions of the ambient space hyperplanes to $Y_4$. On the ambient space, ${\cal A}=\mathbb{P}^1 \times \mathbb{P}^4$, the effective cone is simply the positive quadrant defined by $aH_1 + bH_2$ with $a,b \geq 0$ \cite{Hubsch:1992nu}. However, on $Y_4$ itself there is a richer range of possibilities. For example, the line bundles defined by $L=\cO_{Y_4}(-1, n)$ with $n\geq 5$ all satisfy $h^0(Y_4, L) >0$ and thus their global holomorphic sections define algebraic subvarieties of $Y_4$ even though they cannot be simply described by polynomial defining equations in the ambient coordinates (and $h^0({\cal A}, \cO_{{\cal A}}(-1,n))=0$, $\forall ~n$). In other words, the effective cone of $Y_4$ is larger than simply $a H_1+ b H_2$ with $a,b \geq0$.

It was noted in \cite{Anderson:2015iia} that although divisors of this type are ``non-polynomial" \cite{Green:1987rw,Candelas:1993dm,Candelas:1994hw,Berglund:1994my,Mavlyutov,2004InMat.157..621M} in the homogenous coordinate system of ${\cal A}$, they may still be represented simply as \emph{rational} functions in the ambient space coordinates, that are suitably regular (i.e. holomorphic polynomials) when evaluated on the CY, $Y_4$. To see this explicitly, let us build the global sections of $\cO_{Y_4}(-1, n)$ on $Y_4$. Labeling the homogeneous coordinates of $\mathbb{P}^1, \mathbb{P}^4$ to be $x_i, y_j$ with $i=0,1$ and $j=0, \ldots 4$, consider the defining equation of degree $(2,5)$,
\beq\label{defeqn}
P: x_0^2 {p_5}^{(1)}(y)+x_0 x_1 {p_5}^{(2)}(y)+ x_1^2 {p_5}^{(3)}(y)=0 \;,
\eeq
where ${p_5}^{(a)}(y)$, $a=1,2,3$ are homogeneous quintic polynomials in the $y$-coordinates of $\mathbb{P}^4$. Now, to consider a divisor of the form $\cO_{Y_4}(-1,n)$, by definition it can be decomposed into the associated divisor of zeros and divisor of poles ($D=$(Div. of zeros) - (Div. of Poles)) which in this case can take the form\footnote{In this paper we interchangeably use the following four related terms -- divisor, divisor class, line bundle and global holomorphic section of line bundle -- and freely call one by another unless confusions arise.} 
\beq
\frac{\text{Div. of zeros}}{\text{Div. of poles}}= \frac{f_{n}(y)}{g_{1}(x)} \;.
\eeq
Here $f_n$ is a polynomial of degree $n$ and $g_1$ a polynomial of degree 1.
How then can this rational function be made regular when evaluated on the CY given in \eref{defeqn}? Consider the simple linear function given by $x_0=0$. On this locus, the defining equation $P=0$, guarantees that one specific quintic in the $y$-coordinates also vanishes:
\beq
{p_5}^{(3)}(y)=0 \;.
\eeq
As a result, the rational function 
\beq
\frac{{p_5}^{(3)}(y)}{x_0}~~,
\eeq
is manifestly regular -- every zero of the denominator is matched by a zero of a numerator for points satisfying the defining relation given by \eref{defeqn}. Likewise, by similar logic, $\frac{{p_{5}}^{(1)}(y)}{x_1}$ is also regular and it is straightforward to verify that these two can be used to construct a complete basis of global sections of ${\cal O}_{Y_4}(-1,n)$. To be explicit, a basis of $H^0(Y_4, \cO_{Y_4}(-1,n))$ for $n\geq 5$ is given by
\beq
s= \frac{{p_5}^{(3)}(y)}{x_0} r_{n-5}(y) + \frac{{p_{5}}^{(1)}(y)}{x_1} s_{n-5}(y) \;,
\eeq
where $r_{n-5}(y)$ and $s_{n-5}(y)$ are arbitrary polynomials of degree $(n-5)$ in the $y$-coordinates, each with ${n-1}\choose{4}$ linearly independent monomials, thereby giving rise to a basis for the cohomology with $2{{n-1}\choose{4}}$ elements\footnote{It should be noted that the divisor, $s$, given in this simple illustrative example is in fact singular. All of the  examples that are used in the study of instanton effects in the rest of the paper, however, involve divisors which are smooth.}. \\

\vspace{5pt}
 
The type of construction described in the preceding paragraphs was employed recently in \cite{Anderson:2015iia} to build a new dataset of CY manifolds (including both three-folds and four-folds) which possess a range of interesting features including new Hodge numbers and novel fibration structures. 
In the following sections, the observations above will be employed to find new divisor geometries $D \subset Y_4$ in both finite and infinite families leading to non-trivial superpotentials.

\subsection{A study of instanton superpotentials}
In \cite{Witten:1996bn}, it was argued that the simple form of CICY manifolds make it possible to classify what types of instanton solutions can exist. From remarkably little geometric data, a wide array of conclusions can be reached for several dual string compactifications including M-theory, F-theory, Type IIB and heterotic string theory. In the following sections we will briefly review this structure and point out how the expansion of effective cones described above allows for a broader class of instanton solutions than had been previously explored in \cite{Witten:1996bn} and related work. We will explore the consequences of this for the network of string dualities described above.

In Section \ref{constrandtop} we will provide a brief, self-contained review of the necessary conditions on divisors $D \subset Y_4$ for an $M5$-brane wrapping $D$ to contribute to the $3$-dimensional superpotential and the consequences for dual geometries. In Section \ref{instanton} the necessary conditions for a $5$-brane to contribute to the superpotential, as well as a sufficient condition and a simple topological consistency check are reviewed. In Section \ref{dualities} we review the detailed links between the geometry of $D \subset Y_4$ and the non-perturbative superpotentials in dual F-theory, Type IIB and heterotic vacua. Section \ref{example_sec} provides the first core examples of this work -- both traditional CICY manifolds \cite{Candelas:1987kf,Green:1986ck,Candelas:1987du,Hubsch:1986ny, Gray:2013mja,Gray:2014kda,Gray:2014fla} as well as new gCICY constructions \cite{Anderson:2015iia} -- with the types of (non-``inherited") divisors as described above. In Section \ref{infinite_v_finite} we explore the consequences for the superpotential, in heterotic/F-theory dual pairs, of the situation where $Y_4$ admits a finite or infinite number of divisors capable of leading to instanton contributions to the superpotenial. Finally, in Section \ref{conc} a brief summary and outlook for future work is provided. Appendix \ref{leray} provides useful technical results on fibration structures and line bundle cohomology on CY four-folds.

\section{Instantons in M-theory on CY Four-folds} \label{constrandtop}

\subsection{Instanton geometry}\label{instanton}

We consider M-theory compactified on a smooth CY four-fold, $Y_4$, leading to an $\mathcal{N}=2$ theory in $3$ dimensions. The necessary (but not sufficient) conditions for an M5-brane to contribute non-trivially to the superpotential of the $3D$ theory were clearly laid out in \cite{Witten:1996bn}. To facilitate a self-contained discussion, we will briefly summarize these results here. The first result is that an anomaly computation and consideration of fermion zero modes leads to a necessary condition that must be satisfied in order for a non-trivial superpotential effect to be generated. This can be concisely summarized by the following geometric condition on the arithmetic genus of the holomorphic divisor $D$ on which the $5$-brane is wrapped:
\beq\label{main_cond}
\chi(D, \cO_D)=1 \;.
\eeq
To make sense of this criteria in terms of divisor geometry, we consider the Koszul sequence
\beq\label{koszulD}
0 \to \cO_{Y_4}(-D) \to \cO_{Y_4} \to \cO_D \to 0~.
\eeq
From the fact that $Y_4$ is a CY four-fold and hence, $h^\bullet(Y_4, O_{Y_4})=(1,0,0,0,1)$, the long exact sequence in cohomology associated to \eref{koszulD} yields the following
\begin{subequations}\label{koszuled}
\begin{align}
& h^0(D, \cO_{D})=1-h^0(Y_4, \cO_{Y_4}(-D))+h^1(Y_4,\cO_{Y_4}(-D)) \;,\\
& h^1(D, \cO_{D})=h^2(Y_4,\cO_{Y_4}(-D)) \;,\\
& h^2(D,\cO_{D})=h^3(Y_4,\cO_{Y_4}(-D))\;,  \\
& h^3(D,\cO_{D})=h^4(Y_4,\cO_{Y_4}(-D))-1\;.
\end{align}
\end{subequations}
Defining the index on the four-fold as $\chi(Y_4, \cO_{Y_4}(-D))=\sum_{i=0}^{4} (-1)^i h^i(Y_4, \cO_{Y_4}(-D))$, it is clear that \eref{koszulD} indicates that
\beq
\chi(D, \cO_D)=2- \chi(Y_4, \cO_{Y_4}(-D))~,
\eeq
and therefore,
\beq \label{cond2}
\chi(D, \cO_D)=1 ~~~~~\Leftrightarrow~~~~ \chi(Y_4, \cO_{Y_4}(-D))=1 \;.
\eeq

Within a CY four-fold, this criterion can also be simply written \cite{Donagi:1996yf} in terms of the intersection structure of $D$ inside $Y_4$ as
\beq\label{c2cond}
\chi(D, \cO_D)= - \frac{1}{24}(D^4 + D^2 \cdot c_2(Y_4)) \;,
\eeq
where $D^4$ is the quadruple self-intersection number of $D$ and $c_2(Y_4)$ is the second Chern class of the CY four-fold. Finally, it should be noted that even when \eref{main_cond} is satisfied, if the divisor is not embedded rigidly, the superpotential can vanish due to the presence of additional fermion zero modes or cancellations which can occur when integrating over the $M5$-brane position moduli space \cite{Witten:1996bn} (see also \cite{Beasley:2003fx} and \cite{Braun:2007tp} for similar considerations of cancellations in the context of the heterotic string). In the case of smooth four-folds, these possible cancellations can be avoided if the stronger condition that $D$ has no embedding moduli holds, i.e. 
\beq\label{strong_cond}
h^\bullet(D, {\cal O}_{D})=(1,0,0,0)\;.
\eeq
In terms of cohomologies on $Y_4$, \eref{strong_cond} and \eref{c2cond} lead to
\begin{subequations}\label{cohomDY4}
\begin{align}
& h^0(Y_4,\cO_{Y_4}(D))=1\;,\\
& h^1(Y_4,\cO_{Y_4}(D))=0 \;,\\
& h^2(Y_4,\cO_{Y_4}(D))=0\;, \\
& h^3(Y_4,\cO_{Y_4}(D))=h^4(Y_4,\cO_{Y_4}(D)) = k\; ,
\end{align}
\end{subequations}
where we used Serre duality, $h^i(Y_4, \mathcal O_{Y_4}(-D))=h^{4-i}(Y_4, K_{Y_4}\otimes \mathcal{O}_{Y_4}(D))=h^{4-i}(Y_4, \mathcal{O}_{Y_4}(D))$, and $k$ must equal either $0$ or $1$ (due to the injectivity of the first non-trivial map in the long exact sequence associated with the Koszul sequence,~\eref{koszulD}). In this case no cancellations within the divisor class can take place and we are guaranteed a non-vanishing contribution to the superpotential \cite{Witten:1996bn}.

As a final comment, it is an interesting observation about the structure of the $3$-dimensional effective theory that the number of divisors satisfying \eref{main_cond} may be finite \cite{Witten:1996bn} or infinite \cite{Donagi:1996yf} for a given CY four-fold. In the former case the contributions to the superpotential take a simple form, while the latter case can demonstrate remarkable modular invariance properties (see \cite{Font:1990gx,Ferrara:1989bc,Donagi:1996yf} for discussions). We will explore both types of solution in the following sections.

\subsection{Review of dual geometries}\label{dualities}

One of the main motivations for considering non-perturbative effects in the context of $3$-dimensional compactifications of M-theory is the powerful window such considerations provide into the structure of more phenomenologically relevant $4$-dimensional theories. As first noticed in \cite{Witten:1996bn}, in the case that $Y_4$ admits elliptic or $K3$ fibrations, simple observations about the geometry of the divisor $D \subset Y_4$ yield a variety of information about the structure of the superpotentials in a network of dual $4$-dimensional $\mathcal N=1$ theories.

If $Y_4$ admits a genus-$1$ or elliptic fibration
\beq\label{ell_fib}
\pi: Y_4 \to B_3\;,
\eeq
then it is possible to comment on the superpotentials of the dual $4$-dimensional F-theory and Type IIB vacua. The key distinction in these cases is whether or not the divisor, $D$, is ``vertical" or ``horizontal" with respect to the fibration in \eref{ell_fib}. That is, the distinction is made between the following two possibilities:
\begin{itemize}
\item $D$ is a section or multisection of the elliptic fibration. (``Horizontal").
\item $D$ is the pullback of a divisor on the base $B_3$ (i.e. $D=\pi^{-1}(D_{B_3})$ for some divisor $D_{B_3} \subset B_3$). (``Vertical").
\end{itemize}
It is interesting to note that this key distinction in instanton physics can be made independent of the existence of a section to the fibration in \eref{ell_fib}. The effective physics of F-theory compactified on a genus-$1$ fibered manifold (with multisection) and its associated discrete symmetries has recently become a topic of active investigation (for recent work see e.g. \cite{Braun:2014oya,Morrison:2014era,Anderson:2014yva}). The interplay of such symmetries and non-perturbative physics is an intriguing area of open investigation.

As argued in \cite{Witten:1996bn}, the horizontal divisors only contribute non-trivially to the superpotential in the $3$-dimensional compactification. M-theory on $Y_4$ is dual to Type IIB on $B_3 \times S^1$. If $\epsilon$ is the area of the elliptic fibers of $\pi$, then the volume of a horizontal divisor is a factor of $\epsilon^{-1}$ different to the volume of a vertical divisor, with $\epsilon \to 0$ being the Type IIB/F-theory limit. As a result, a simple scaling argument shows that contributions from horizontal divisors vanish in the $\epsilon \to 0$ limit. It should be noted that contributions from both horizontal and vertical divisors may appear when one considers other, generically strongly coupled, regimes of the theory. The complete knowledge of these superpotential contributions can be also useful in the context of strong-weak dualities.

The second class, of ``vertical" divisors pulled back from the base\footnote{Note, we will refer to a divisor as vertical even if it only contains vertical components.}, can lead to $D3$ brane instanton contributions to the 4-dimensional effective theory in Type IIB/F-theories. Depending on the structure of singular fibers of $Y_4$, such divisors can be either reducible or singular \cite{Braun:2000hh}. For the present consideration, however, we will restrict ourselves to the case that $D$ is smooth and irreducible (for instance in the case that all fibers of $Y_4$ are irreducible).

In the case of vertical divisors it is clear that $D^4=0$ and thus, the topological check given in \eref{c2cond} takes the simple form 
\beq\label{c2again}
\chi(\cO_D)=-\frac{1}{24}D^2 \cdot c_2 (Y_4)\;.
\eeq
In \cite{Donagi:1996yf} it was observed that $D^{n-2} \cdot c_2 \geq 0$ for all nef divisors in an $n$-dimensional CY manifold \cite{yau,miyaoka}. Thus, it is clear from \eref{c2again} that any vertical divisors satisfying \eref{main_cond} must be non-nef.

In the case that the theory also admits a $K3$ fibration (suitably compatible with the elliptic fibration described above), we can also comment on the dual heterotic theory \cite{Friedman:1997yq}. In particular, heterotic/F-theory duality requires that the geometries form a pair
\beq\label{het_f}
\text{Heterotic on}~~\pi_h: X_3 \stackrel{\mathbb{E}}{\longrightarrow} B_{2}~~~~ \Leftrightarrow~~~~ \text{F-theory on}~~ \rho_f: Y_{4} \stackrel{K3}{\longrightarrow} B_{2}\;,
\eeq
where the fibrations are compatible in that they share a common base, and the base $B_3$ of the elliptic fibration~\eref{ell_fib} is itself rationally fibered over $B_2$ via
\beq
\tau: B_3 \stackrel{\mathbb{P}^1}{\longrightarrow} B_{2} \ . 
\eeq
In the case that the divisor, $D$, is not a section or multisection of the elliptic fibration of $Y_4$, it non-trivially contributes to the superpotential of the $\mathcal N=1$, $4$-dimensional F-theory EFT and we would expect this to also lead to contributions in the heterotic theory. In \cite{Witten:1996bn}, these contributions were distinguished with respect to their projection under the rational fibration, $\tau$:
\begin{itemize}
\item $D_{B_3} \subset B_3$ is a section of the $\mathbb{P}^1$ fibration $\tau(D_{B_3}) \simeq B_2$ (``$\tau$-Horizontal").
\item $\tau(D_{B_3}) \subset B_2$ (``$\tau$-Vertical").
\end{itemize}
The first of these cases corresponds to spacetime instanton contributions to the $\mathcal N=1$, $4$-dimensional heterotic superpotential, while the second leads to world sheet instanton contributions. These basic duality results are summarized in Table \ref{table1}.  For a further discussion on the dualities between NS5-brane solutions in heterotic theories wrapping the elliptic fibers of $X_3$ or divisors $D_{B_2} \subset B_2$ see \cite{Friedman:1997yq,Diaconescu:1999it}.
\begin{table}
\begin{centering}
\begin{tabular}{|c||c|c|c|c|}\hline
 & M-theory & F-theory & IIB & Heterotic \\ \hline
 Section (or multisection) of $\pi$ & \checkmark & $\times$  & $\times$ & $\times$ \\ \hline
$D$ contains $\pi^{-1}(D_{B_3})$ & \checkmark & \checkmark & \checkmark & \checkmark \\ \hline
 $\tau(D_{B_3}) \simeq B_2$ ~ $(D \sim \pi^{-1}(D_{B_3}))$ & \checkmark & \checkmark & \checkmark & \checkmark (SPI) \\ \hline
$\tau(D_{B_3}) \subset B_2$ ~ $(D \sim \pi^{-1}(D_{B_3}))$& \checkmark & \checkmark & \checkmark & \checkmark (WSI) \\ \hline
 \end{tabular}
\caption{\it Description of trivial ($\times$) vs. non-trivial ($\checkmark$) superpotential contributions in different dual theories. Here we assume that a divisor $D \subset Y_4$ satisfies $h^\bullet(D, {\cal O}_{D})=(1,0,0,0)$ in M-theory on a CY four-fold $Y_4$ which admits an elliptic fibration ($\pi: Y_4 \to B_3$). In the case that in addition $Y_4$ admits a compatible $K3$ fibration such that $\tau: B_3 \to B_2$ (with $\mathbb{P}^1$ fiber), a dual heterotic theory also exists. In the heterotic theory ``WSI " refers to a world sheet instanton and ``SPI" to a spacetime instanton. }\label{table1}
\end{centering}
\end{table}

To conclude, it is useful to make one more important distinction in the case of torus fibered CY four-folds. The existence of a ``horizontal" divisor (i.e. a section or multisection) does not automatically guarantee the existence of a divisor $D$ with arithmetic genus one. Instead, it should be noted that \emph{sections} (either holomorphic or rational) must contribute non-trivially to the superpotential, while multi-sections are not necessarily even of the correct arithmetic genus.

First, we consider so-called ``holomorphic" and ``rational" sections (see for example, \cite{Morrison:2012ei,Mayrhofer:2012zy,Cvetic:2013nia}). A holomorphic section defines the base, $B_3$, as a sub-variety of $Y_4$ and moreover can be expressed as a holomorphic (polynomial) function of the base coordinates. On the other hand, ``rational" sections define a sub-variety $\tilde{B}_3 \subset Y_4$ which is \emph{birational} to $B_3$. Both holomorphic and rational sections automatically satisfy the stronger condition \eref{strong_cond}. This can be seen simply from the Koszul sequence:
\beq
0 \to \cO_{Y_4}(-S) \to \cO_{Y_4} \to \cO_{S} \to 0 \;.
\eeq
Since $S_{hol}=0$ defines the base $B_3$ as an algebraic subvariety of $Y_4$, it is clear that if $h^i(S_{hol},\cO_{S_{hol}})$ $\neq 0$, for $i=1,2,3$, then the holomorphic $i$-forms would pull back non-trivially to $Y_4$ under the projection map $\pi: Y_4 \to B_3$, in contradiction to the CY condition. As a result, any holomorphic section must in fact have not only arithmetic genus equal to 1 but also satisfy the stronger condition that $h^\bullet(S_{hol},\cO_{S_{hol}})=(1,0,0,0)$. On the other hand, a rational section is only birational to $B_3$ (and can ``wrap" non-trivial blow-up directions in the (resolution) of the elliptic fiber). As a result, its bundle-valued cohomology could in principle differ from that of $B_3$. However, since $h^r(B,\cO_B)$ is a birational invariant \cite{kollar}, here too, we see that if $S_{rational}$ defines a three-fold surface, $\tilde{B}_3$, birational to $B_3$ inside of $Y_4$, it will also satisfy \eref{strong_cond}.

In contrast, it should also be noted here that multisections \cite{Braun:2014oya,Morrison:2014era} do not generically have arithmetic genus equal to $1$. For instance,
\begin{eqnarray} \label{eg4}
Y_4 = \left[ \begin{array}{c||c c} 
\mathbb{P}^1 & 2  \\ 
\mathbb{P}^1 & 2 \\ 
\mathbb{P}^1 & 2 \\
\mathbb{P}^1 & 2 \\
\mathbb{P}^1 & 2 \\
\end{array}\right]~,
\end{eqnarray}
is a genus-1 fibered CY four-fold which does not possess a section, but instead only multisections of order $2$ at best (e.g. $\cO(1,0,0,0,0)$), all with vanishing arithmetic genus. Many examples of torus-fibered CICY or gCICY CY manifolds in fact have \emph{only} multisections and in such cases, are not guaranteed to give rise to any superpotential terms even in the M-theory limit.

\section{Examples}\label{example_sec}

\subsection{Example 1: A CICY four-fold} \label{constr}
In this section we demonstrate that smooth CICY four-folds as constructed in \cite{Brunner:1996bu} and fully classified in \cite{Gray:2013mja,Gray:2014kda,Gray:2014fla} can admit divisors with arithmetic genus $1$ (and indeed rigid divisors). As one simple illustration of this, we will consider an elliptically and $K3$-fibered manifold which admits a section to its elliptic fibration. The following CY four-fold
\begin{eqnarray} \label{eg1}
Y_4 = \left[ \begin{array}{c||ccccc} 
\mathbb{P}^1 & 1 & 1 & 0 & 0 & 0 \\ 
\mathbb{P}^2 & 0 & 1 & 0 & 1 & 1 \\ 
\mathbb{P}^3 & 1 & 0 & 1& 1 & 1 \\
\mathbb{P}^1 & 1 & 0 & 1&  0 & 0 \\
\mathbb{P}^1 & 1 & 0 & 0&  1 & 0 \\
\mathbb{P}^1 & 0 & 0 & 1&  0 & 1 \\
\end{array}\right]\;~, 
\end{eqnarray}
has Euler number $\chi(Y_4)=456$ and the Hodge numbers, 
$h^{1,1}=6, \,\, h^{3,1}=62, \,\,h^{2,2}=316$. 

This manifold is elliptically fibered over $\mathbb{P}^1 \times \mathbb{P}^1 \times \mathbb{P}^1$, $\pi: Y_4 \stackrel{\mathbb{E}}{\rightarrow} \mathbb{P}^1 \times \mathbb{P}^1 \times \mathbb{P}^1$ and has a compatible $K3$-fibration over $\mathbb{P}^1 \times \mathbb{P}^1$, $\rho_f: Y_4 \stackrel{K3}{\rightarrow} \mathbb{P}^1 \times \mathbb{P}^1$. More precisely, the $K3$-fiber is described via the complete intersection:
\begin{eqnarray} \label{eg_k3fib}
K3 = \left[ \begin{array}{c||ccccc} 
\mathbb{P}^1 & 1 & 1 & 0 & 0 & 0 \\ 
\mathbb{P}^2 & 0 & 1 & 0 & 1 & 1 \\ 
\mathbb{P}^3 & 1 & 0 & 1& 1 & 1 \\
\mathbb{P}^1 & 1 & 0 & 1&  0 & 0 \\
\end{array}\right]\;,
\end{eqnarray}
which in turn is elliptically fibered over $\mathbb{P}^1$ with fiber
\begin{eqnarray} \label{eg_efib}
\mathbb{E} = \left[ \begin{array}{c||ccccc}
 \mathbb{P}^1 & 1 & 1 & 0 & 0 & 0 \\ 
\mathbb{P}^2 & 0 & 1 & 0 & 1 & 1 \\ 
\mathbb{P}^3 & 1 & 0 & 1& 1 & 1 \\
\end{array}\right]\;.
\end{eqnarray}
On the four-fold given in \eref{eg1}, let $H_i$, with $i=1,\ldots 6$, denote the divisors obtained by restriction of the ambient projective space factor hyperplanes. Then, in this notation, the divisor
\beq\label{sec1}
-H_1 + H_2
\eeq
is in fact a section to the elliptic fibration described above. Using the tools described in the Introduction, the global sections of $\cO_{Y_4}(D)$ associated to this effective divisor can be described as follows.

Denote the homogeneous coordinates of the six projective space factors by $ {\bf x} \in \IP^1,  {\bf y} \in \IP^2,  {\bf z} \in \IP^3 ,  {\bf u} \in \IP^1,  {\bf v} \in \IP^1,  {\bf w} \in \IP^1$ respectively. The five defining relations of the complete intersection can be written as $P^i({\bf x, y, z, u, v, w})$, with $i=1,2,\dots, 5$. Explicitly, $P^2$ and $P^3$ for example, take the following form
 \begin{eqnarray}
P^2({\bf x, y}) &=& x_0 \, p^{2  \, (1)}_{1} ({\bf y}) + x_1\, p^{2 \, (2)}_{1} ({\bf y}) \\
P^3({\bf z,u,w}) &=& u_0\, p^{3 \,(1)}_{11} ({\bf z, w)} + u_1\, p^{3\, (2)}_{11}({\bf z, w}) \nonumber ~,
 \end{eqnarray}
 where $p^{2(i)}_{1}$ are linear functions in ${\bf y}$ and $p^{3(j)}_{11}$ are multi-degree $(1,1)$ in ${\bf (z,w)}$. The divisor can be described uniquely, up to an overall factor, as: 
 \bea
 \frac{p^{2 (2)}_{1} ({\bf y})}{ x_0} =0\;.
 \eea
It can be verified that
\beq
h^\bullet(Y_4, \cO_{Y_4}(1,-1,0,0,0,0))=(0,0,0,0,1) \;,
\eeq
so the cohomology of ${\cal O}_{Y_4}(-D)$, satisfies the criteria laid out in \eref{cond2}, and the stronger one given by \eref{strong_cond}, and \eref{cohomDY4} with $k=0$. Since this divisor is, by construction, a section of the fibration, $D \subset Y_4$ is a copy of the entire $\mathbb{P}^1 \times \mathbb{P}^1 \times \mathbb{P}^1$ base.

As described above this divisor provides an instanton superpotential in M-theory which provides a trivial contribution when dualized into F-theory/Type IIB. However, the four-fold given in \eref{eg1} is also $K3$ fibered and as a result, we can consider the heterotic dual theory as well. 

To generate non-trivial instanton contributions to F-theory, and thus heterotic string theory, on \eref{eg1}, one can consider a second example divisor
\beq
\cO_{Y_4}(0,0,1,-1,0,1)\;.
\eeq
The global sections of this line bundle can be explicitly realized, in a similar manner to the case described above, as
  \bea
 \frac{p^{3 (2)}_{11} ({\bf z, w})}{ u_0}\;,
 \eea
which is again unique up to an overall factor. Once again $h^\bullet(D, \cO_D)=(1,0,0,0)$, where \eref{cohomDY4} is satisfied with $k=0$, and $\chi(D,\cO_D)=1$. But here we get a non-trivial superpotential contribution in F-theory generated by $D3$ branes wrapping $\pi(D)$. These dualize in the heterotic theory into world sheet instanton contributions. 

\subsection{Example 2: A gCICY four-fold}

In this section we explore another example, a gCICY manifold, using the tools described in Section \ref{first_tools}. A smooth CY four-fold with Euler number $\chi(Y_4)=480$ can be defined by the configuration matrix
\begin{eqnarray} \label{eg2}
Y_4 = \left[ \begin{array}{c||c | c} 
\mathbb{P}^3 & 1 & 3  \\ 
\mathbb{P}^1 & 1 & 1 \\ 
\mathbb{P}^1 & 3 & -1 \\
\mathbb{P}^1 & 1 & 1 \\
\end{array}\right]\; .
\end{eqnarray}
This, according to the gCICY notation \cite{Anderson:2015iia}, characterizes $Y_4$ as a hypersurface inside
\begin{eqnarray} \label{eg2m}
{\cal M}  = \left[ \begin{array}{c||c } 
\mathbb{P}^3 & 1   \\ 
\mathbb{P}^1 & 1  \\ 
\mathbb{P}^1 & 3  \\
\mathbb{P}^1 & 1  \\
\end{array}\right]\;,
\end{eqnarray}
given by a global section of $\cO_\cM(3,1,-1,1)$.
The independent Hodge numbers of this four-fold are given by $h^{1,1}=4,~h^{3,1}=68,~h^{2,2}=332$.
Denoting the homogeneous coordinates of the four ambient projective space factors in turn by ${\bf x}=(x_0 :  x_1 : x_2 : x_3)$, ${\bf y}=(y_0 :  y_1)$, ${\bf z}=(z_0 :  z_1)$ and ${\bf u}=(u_0 :  u_1)$, the defining equation for $\cM$ can be written as
\begin{eqnarray}
\label{eq:first}
\hspace{-0.5cm} P({\bf x,y,z,u})= z_0^3\, p_{111}^{(1)} ({\bf x,y,u}) + z_0^2 z_1\, p_{111}^{(2)} ({\bf x,y,u}) + z_0 z_1^2 \,p_{111}^{(3)} ({\bf x,y,u}) + z_1^3 \,p_{111}^{(4)} ({\bf x,y,u})
\end{eqnarray}
Here, the  $p_{111}^{(a)} ({\bf x,y,u})$, where $a=1,\ldots,4$, are generic homogeneous tri-linear polynomials.
To obtain the explicit expression for a section $Q \in H^0(\cM, \cO_\cM(3,1,-1,1))$ that defines the embedding of $Y_4$ in $\cM$, we follow the gCICY construction method~\cite{Anderson:2015iia}. According to the degree-splitting rule there, $Q$ is taken to have the rational form,
\begin{equation}\label{Q}
Q= \frac{f({\bf x,y,u})}{g({\bf z})}\ , 
\end{equation}
where $f$ and $g$ are, respectively, polynomials of multi-degree $(3,1,0,1)$ and $(0,0,1,0)$ in the $\bold x$, $\bold y$, $\bold z$, and $\bold u$ coordinates. Now, with the denominator choice of $g({\bf z})= z_0$, the corresponding numerator polynomial $f({\bf x,y,u})$  should vanish on the divisor $z_0=0$ of $\cM$. On the other hand, the defining equation for $\cM$, (\ref{eq:first}), also vanishes on $\cM$ by construction, which indicates that $p_{111}^{(4)} ({\bf x,y,u}) =0$ on the locus $z_0=0$ inside $\cM$. Given this, we can obtain an appropriate numerator by multiplying $p_{111}^{(4)} ({\bf x,y,u})$ with a quadric polynomial in $\bold x$, thus ending up with the following 10 global sections of line bundle ${\cal O_M} (3,1,-1,1)$:
\begin{equation} \label{onelab}
s_i = \frac{p_{111}^{(4)} ({\bf x,y,u})}{z_0} m_{2,i}(\bold x)\quad\quad i=1,\dots,10~.
\end{equation}
Here $m_{2,i}(\bold x)$, for $i=1, \cdots, 10$ are the ten quadratic monomials, $(x_0)^2, x_0\, x_1, \cdots, (x_3)^2$, in the coordinates $\bold x = (x_0: x_1:x_2:x_3)$.
One can also choose different  denominators, such as $g({\bf z})=z_0-z_1$ and $g({\bf z})=z_0+z_1$, which, respectively, give rise to 10 more sections each:
\begin{subequations} \label{twolab}
\begin{eqnarray} 
t_i &=& \frac{\sum\limits_{a}^4 p_{111}^{(a)} ({\bf x,y,u})}{z_0-z_1} m_{2,i}(\bold x)  
\quad\quad i=1,\dots,10~, \\
u_i &=& \frac{\sum\limits_{a}^4 (-1)^a p_{111}^{(a)} ({\bf x,y,u})}{z_0+z_1} m_{2,i}(\bold x)  
\quad\quad i=1,\dots,10~.
\end{eqnarray}
\end{subequations}
It is easy to verify that these 30 sections are linearly independent and that any other section constructed by choosing a different denominator from $z_0, z_0-z_1, z_0+z_1$ can be written as a linear combination of (\ref{onelab}) and (\ref{twolab}).
Given that $h^0({\cal M}, {\cal O_M} (3,1,-1,1))=30$, we conclude that $s_i$, $t_i$, and $u_i$ span the entire section space $H^0(\cM, {\cal O_M} (3,1,-1,1))$.   As a result, the defining equation, $Q$, for $Y_4 \subset \cM$, can be written as
\begin{eqnarray}
Q= \sum\limits_{i=1}^{10} \,\alpha_i \, s_i + \sum\limits_{i=1}^{10} \,\beta_i \, t_i +\sum\limits_{i=1}^{10}  \, \gamma_i\,u_i,
\end{eqnarray}
where $\alpha_i, \beta_i$, and $\gamma_i$, for $i=1,\dots, 10$, are  generic complex coefficients. 
One can further check that the resulting gCICY, $Y_4$, is smooth for a generic choice of $P$ and $Q$. 

Once again this manifold is torus fibered ($\pi: Y_4 \stackrel{\mathbb{E}}{\longrightarrow} \mathbb{P}^1 \times \mathbb{P}^1 \times \mathbb{P}^1$) and $K3$-fibered ($\rho_f: Y_4 \stackrel{K3}{\longrightarrow} \mathbb{P}^1 \times \mathbb{P}^1$). Note that the torus fibration of the manifold does not necessarily admit a section in this case. Here the $K3$ fiber is given by
\begin{eqnarray} \label{eg2_k3}
K3= \left[ \begin{array}{c||c c} 
\mathbb{P}^3 & 1 & 3  \\ 
\mathbb{P}^1 & 1 & 1 \\ 
\end{array}\right]\;,
\end{eqnarray}
which is in turn torus fibered,
\begin{eqnarray} \label{eg2_elliptic}
T^2= \left[ \begin{array}{c||c  c} 
\mathbb{P}^3 & 1 & 3  \\ 
\end{array}\right] (\simeq \left[ \mathbb{P}^2 | | 3 \right] \text{over any given point in the base})\;.
\end{eqnarray}

In this example we find a ``vertical" instanton which leads to a non-trivial superpotential contribution in F-theory according to the distinction made in Section \ref{constrandtop}. Taking
\beq
D \sim \cO_{Y_4}(1,-1,3,1)\;,
\eeq
it can be verified using the techniques developed in \cite{Anderson:2015iia} that $h^\bullet(D, \cO_D)=(1,0,0,0)$. Moreover this divisor is \emph{not} a section of the fibration, but also includes non-trivial base dependence. Since the geometry admits both $K3$ and $T^2$ fibrations (and they are compatible), the consequences of this divisor for dual theories are readily ascertainable. As shown in Table \ref{table1}, $D$ above will lead to a non-trivial instanton superpotential not only in M-theory, but also in the dual F-theory/Type IIB theories. Moreover, since this geometry is also $K3$ fibered, we see that 
$\rho(D)$ is a non-trivial curve in the two-fold base -- $\mathbb{P}^1 \times \mathbb{P}^1$ -- of the heterotic dual (i.e. it is also not a section to the $K3$-fibration). As a result this divisor leads to a non-trivial world-sheet instanton effect in the $4$-dimensional heterotic dual theory.

With these examples in hand, we turn now to a more systematic study of instantons in dual heterotic/F-theory effective theories.

\section{Heterotic/F-theory Dual Pairs and Finite vs. Infinite Families of Solutions}\label{infinite_v_finite}
In this section we turn our attention to two important questions:
\begin{itemize}
\item Under what conditions is it possible for an \emph{infinite family} of divisors to contribute to the superpotential?
\item Is it possible to characterize divisors (and potentially infinite families as above) that will contribute to the superpotential in heterotic/F-theory dual pairs?
\end{itemize}
Beginning with the first point above, an important distinction can be made between four-fold geometries which admit only a finite number of divisors with arithmetic genus $1$ (i.e. \eref{main_cond}) and those that admit infinitely many such divisors. In the latter case, the structure of the superpotential can exhibit interesting modular behavior (see \cite{Font:1990gx, Ferrara:1989bc} for early conjectures and \cite{Donagi:1996yf} for an explicit modular superpotential with $E_8$ symmetry). We start by observing that it is straightforward to engineer examples of CICY (or gCICY) four-fold geometries with an infinite family of divisor classes with arithmetic genus $1$. 

\subsection{An infinite family of divisors with arithmetic genus $1$}
Consider the following CY four-fold,
\begin{eqnarray} \label{eg_inf}
Y_4= \left[ \begin{array}{c||cccc} 
\mathbb{P}^1 & 0 & 0 & 1 & 1 \\ 
\mathbb{P}^1 & 0 & 1 & 0 & 1 \\ 
\mathbb{P}^1 & 1 & 0 & 1 & 0 \\ 
\mathbb{P}^1 & 0 & 0 & 0 & 2 \\ 
\mathbb{P}^1 & 0 & 0 & 2 & 0 \\ 
\mathbb{P}^3 & 1 & 1 & 2 & 0 \\ 
\end{array}\right]\;,
\end{eqnarray}
and the family of divisors
\beq\label{infclass}
D_a \sim \cO_{Y_4} (a, -1, a, 0,0,1) \ ,  \quad a \geq 0
\eeq
parametrized by integer $a$. The cohomology computation leads to,
\beq
h^\bullet(Y_4, \cO_{Y_4}(D_a))= ((a+1)^2 , a^2+2a, 0,0,0) \ , 
\eeq
from which we see that the divisor $D_a$ is effective and the Koszul sequence \eref{koszulD} leads to 
\beq\label{infclasscohom}
h^\bullet(D_a, \cO_{D_a})=(1, 0, a^2+2a, a^2 + 2a) \ , 
\eeq
which in particular gives $\chi(D_a, \cO_{D_a})=1$. 
Therefore, for $a=0$, the cohomology \eref{infclasscohom} guarantees a non-trivial superpotential contribution, while each of the divisors with $a\neq 0$ in \eref{infclass} is a potential source for a non-trivial superpotential term. Since these divisors are not rigidly embedded, further analysis is required to determine whether each member of the family survives possible cancellations to contribute to the superpotential. Because of such possible cancellations, it is difficult to directly analyze the structure of the superpotential and any possible modular behavior. As a result, it is intriguing to search for infinite families satisfying the stronger condition in \eref{strong_cond}. In the majority of the literature (see \cite{Donagi:1996yf}), M-theory superpotentials with modular symmetry involve heterotic/F-theory dual pairs and divisors of a very special form. To explore this we turn now to heterotic/F-theory dual pairs and infinite families within this context.

\subsection{Instanton families in Heterotic/F-theory dual pairs}
In the context of heterotic/F-theory duality, one particularly rich class of instanton solutions for an elliptically/$K3$ fibered CY four-fold includes divisors that are pulled back from the two-fold base, $B_2$ in \eref{het_f}. If divisors are found in this class with arithmetic genus equal to one, or the stronger, rigidly embedded condition in \eref{strong_cond}, this provides insight into the non-perturbative superpotential of both the dual $\mathcal N=1$, $4$-dimensional theories. In this case, a relevant divisor $D_{B_2} \subset B_2$ will also pullback non-trivially to a divisor in the elliptically fibered three-fold, $X_3$ in \eref{het_f} and can lead to a world sheet instanton contribution to the heterotic superpotential.

It should be noted that the consideration of such dual pairs is particularly interesting due to the involved structure of the moduli dependent prefactors which appear in the superpotential. In heterotic effective theory, one must consider not only the isolated/rigid curves that contribute to the superpotential, but also the bundle-moduli dependent Pfaffian factors (which vanish if the bundle restricts non-trivially to the curve in $B_2$) \cite{Buchbinder:2002ic,Buchbinder:2002pr,Beasley:2003fx,Beasley:2005iu,Braun:2007xh,Braun:2007tp,Braun:2007vy}. Such calculations can be compared with analogous computations on the M-/F-theory side \cite{Cvetic:2012ts} to yield non-trivial support for both methodologies. Important information to obtain before such considerations, however, is a systematic consideration of the divisors (pulled back from $B_2$) in $Y_4$ \cite{Donagi:1996yf,Curio:1997rn,Diaconescu:1999it,Cvetic:2012ts}. In the following paragraphs, we will systematically explore several solutions of this type and consider under what conditions we can find heterotic/F-theory dual theories with superpotentials exhibiting modular behavior.

\subsubsection{Pulling back divisors from $B_2$} \label{sssec:pullB2}
Suppose that there exists a divisor $D_{B_2} \subset B_2$ with cohomology $h^{\bullet}(B_2, \cO_{B_2}(D_{B_2}))=(1,0,0)$. Under what conditions will this pull back to a divisor of $Y_4$ with arithmetic genus equal to one and satisfying the strong condition, $h^{\bullet}(Y_4, \rho^{*}(\cO_{B_2}(D_{B_2})))=(1,0,0,k,k)$, of \eref{cohomDY4}?

A simple tool to answer this question is provided by the Leray spectral sequence for bundle-valued cohomology on a fibered manifold (see \cite{shafarevich} for a review). As discussed in Appendix \ref{leray}, we find the following criteria,
\begin{align}\label{cond_needed}
& h^\bullet(B_2, \cO_{B_2}(D_{B_2}))=(1,0,0)  \ ,\\
& h^\bullet(B_2, \cO_{B_2}(D_{B_2}) \otimes K_{B_2})=(0,k,k) \ ,\nonumber
\end{align}
to guarantee that a (rational) curve in $B_2$ will pull back to a rigidly embedded divisor  in $Y_4$ with $h^\bullet (D, \cO_D)=(1,0,0,0)$.
Moreover, it is known that a curve in $B_2$ satisfying the above criteria is an isolated, rational curve in $B_2$ (i.e. a curve of genus zero).

To see this, recall that by the Riemann-Roch theorem \cite{AG}, the genus of a curve $C \subset B_2$ is given by
\beq
2g -2 =C \cdot (C+K_{B_2}) \ ,
\eeq
while the Euler characteristic (i.e. index) of any smooth curve is in turn:
\beq
\sum_i (-1)^i h^i(B_2, \cO_{B_2}(C))=\chi(B_2, \cO_{B_2}(C))=\frac{1}{2} C \cdot (C-K_{B_2})+\chi(B_2, \cO_{B_2}) \ .
\eeq
Furthermore, for the base $B_2$ of a torus-fibered CY three-fold (the heterotic geometry) it is clear that $\chi(B_2, \cO_{B_2})=1$. Thus, combining the formulae above with the required conditions in \eref{cond_needed}, it is clear that the index of $D_{B_2} + K_{B_2}$ is
\beq
\chi(B_2, \cO_{B_2}(D_{B_2} + K_{B_2}))=0=\frac{1}{2} (D_{B_2}+ K_{B_2}) \cdot (D_{B_2})+1=g-1+1 \ ,
\eeq
hence, $g=0$. Thus, as expected, a search for divisors in $B_2$ contributing to the superpotential leads to a consideration of rational curves. Since a rational curve in $B_2$ always obeys the proposed criteria~\eref{cond_needed}, the divisors of $B_2$ with these properties have to be in one-to-one correspondence with the rational curves in $B_2$. We can now ask, for what complex surfaces, $B_2$ can we expect an infinite number of genus zero curves, $D_{B_2}$?

To begin, it should be observed that the work of Grassi \cite{grassi}, Gross \cite{gross} and the minimal model program \cite{Reid,Barth} has led to a characterization of the possible surfaces, $S$, which can support an elliptically (or genus-$1$) fibered CY three-fold. This set consists of the following surfaces, $\mathbb{P}^2$, the Enriques surface, the Hirzebruch surfaces $\mathbb{F}_m$, for $0 \leq m \leq 12$, and the blow-ups of these surfaces at one or more points. A systematic approach towards enumerating and classifying these non-minimal (i.e. blown-up) surfaces has recently been undertaken in \cite{Morrison:2012np,Morrison:2012js,Martini:2014iza,Johnson:2014xpa,Taylor:2015isa,Taylor:2015ppa} and has led to a dataset of tens of thousands of distinct toric surfaces (see \cite{Morrison:2012np,Morrison:2012js}) and some non-toric geometries (with $h^{1,1}(S) < 8$) \cite{Taylor:2015isa}.

These surfaces then form the arena for our question: How many of them can admit infinite families of rational curves? Of the minimal set, $\mathbb{P}^2$ and $\mathbb{F}_m$, as well as the del Pezzo surfaces, $dP_r$ with $0 \leq r \leq 8$ can be immediately ruled out, as all are known to contain only finitely many rational curves. More interesting are the family of surfaces including $K3$, the Enriques surfaces, and the rationally elliptically fibered surface ($dP_9$) \cite{Donagi:1996yf}. Each of these admits an elliptic fibration over $\mathbb{P}^1$
\beq
\pi_{S}: S \to \mathbb{P}^1~,
\eeq
with sections $\sigma_{i}$. It is the presence of more than one such section in these cases -- that is, a non-trivial Mordell-Weil group -- which generates an infinite family of sections and hence, of rational curves (since each holomorphic section to the elliptic fibration is a copy of the $\mathbb{P}^1$ base). In the case of $dP_9$, the Mordell-Weil group is famously large (rank $8$) and leads to a space of sections (with self-intersection $C^2=-1$) which are linked to the root lattice of $E_8$ \cite{oguiso_shioda}. If $dP_9$ forms the base of a $K3$-fibered CY four-fold, it is natural to consider the pull-back of infinite families of such sections as divisors in $Y_4$. The contribution of these Mordell-Weil elements to the superpotential was studied in \cite{Donagi:1996yf,Curio:1997rn} and found to lead to a remarkable $E_8$ modular symmetry. Since the $K3$ surface cannot serve as base to a non-trivial CY$_3$ elliptic fibration and the Enriques surfaces leads to an essentially trivial Weierstrass model, the $dP_9$ surface remains one of the most interesting examples which we will explore in detail below.

To conclude, we consider how the existing datasets of bases $B_2$ available could be explored for infinite families and modular superpotentials in the future. A result due to Bogomolov (see \cite{deschamps,miyaoka2}) states that if $S$ is a surface of general type with
\beq\label{rational_test}
c_{1}^2(S) > c_2(S) \ ,
\eeq
then for any $g$, the curves of geometric genus $g$ on $S$ form a bounded family. Since a surface of general type cannot be covered by rational curves, these curves cannot deform. So this result implies that a surface $S$ satisfying \eref{rational_test} contains only finitely many rational curves. This criteria could be employed to filter the dataset of surfaces $B_2$ for those leading to infinite families. Some surfaces constructed already in \cite{Martini:2014iza} are similar to $dP_9$ in that they are known to contain $(-1)$-curves and possess infinitely generated Mori cones. It would be interesting to explore the possible modular structure of such examples in the future. For now, we simply return to the $dP_9$ surface to illustrate that the techniques developed in this work readily lead to infinite, modular families. In this case we find a class of instanton contributions leading to an $SU(2)$ symmetry in the superpotential.

\subsubsection{An infinite family on $dP_9$}
In this section we will study a fibration $\rho_f: Y_4 \stackrel{K3}{\rightarrow} dP_9$ similar to that investigated in \cite{Donagi:1996yf,Curio:1997rn}. Consider the following CY four-fold,
\begin{eqnarray} \label{eg2_inf}
Y_4= \left[ \begin{array}{c||ccccc} 
\mathbb{P}^2 & 0 & 0 & 3  \\ 
\mathbb{P}^1 & 0 & 0 & 2  \\ 
\mathbb{P}^1 & 1 & 1 & 0  \\ 
\mathbb{P}^2 & 1 & 2 & 0 \\ 
\mathbb{P}^1 & 1 & 0 & 1  \\ 
\end{array}\right]\;,
\end{eqnarray}
and the family of divisors,
\beq\label{infclass2}
D_a = \cO_{Y_4} (0,0, -1 + 3 a, 1 - 2 a, 1 - 8 a + 10 a^2) \ , \quad a \in \mathbb Z \ .
\eeq
We claim that each divisor $D_a$ satisfies 
\beq\label{infclasscohom2}
h^\bullet(D_a, \cO_{D_a})=(1, 0, 0,0) \ ,
\eeq
and, therefore, a non-trivial superpotential term is generated by each $D_a$ in the family~\eref{infclass2} (for the \emph{smooth} four-fold above). 
To see how \eref{infclasscohom2} comes about, note first that the four-fold geometry is $K3$-fibered, 
\beq
\rho: Y_4 \to B_2 \ ,
\eeq
where the base $B_2$ has the following configuration matrix,
\begin{eqnarray} \label{eg2_inf_base}
B_2= \left[ \begin{array}{c||cc} 
\mathbb{P}^1 & 1 & 1   \\ 
\mathbb{P}^2 & 1 & 2  \\ 
\mathbb{P}^1 & 1 & 0   \\ 
\end{array}\right]\;,
\end{eqnarray}
which describes a $dP_9$ surface. Then, the cohomology in~\eref{infclasscohom2} originates from the following observations on the base $B_2$,
\begin{eqnarray}
&&h^\bullet(B_2,\; L_a)=(1, 0, 0) \label{thingy1}\ , \\
&&h^\bullet(B_2,\; K_{B_2} \otimes L_a)=(0,0,0) \nonumber\ , 
\end{eqnarray}
where 
\beq
L_a=\cO_{B_2} (-1+3a, 1-2a, 1-8a+10a^2)
\eeq
satisfies $D_a = \pi^*{L_a}$ and $K_{B_2} = \cO_{B_2} (0,0,-1)$ is the canonical bundle of $B_2$. This family of divisors are in fact holomorphic sections of the elliptic fibration visible in \eref{eg2_inf_base}. The cohomology group in \eref{thingy1} on the base can be pulled back to $Y_4$ via the Leray spectral sequence (see Appendix~\ref{lerayK3} for details)
\begin{eqnarray}
H^0(Y_4, D_a)&\simeq& H^0(B_2, L_a) = \IC \nonumber\ , \\
H^1(Y_4, D_a) &\simeq&H^1(B_2, L_a) = 0 \ , \\
H^i(Y_4, D_a) &\simeq& H^{i}(B_2, L_a) \oplus H^{i-2} (B_2, L_a \otimes K_{B_2}) =0 \ , \quad i\geq 2 \ \nonumber ,
\end{eqnarray}
which then leads to the desired result~\eref{infclasscohom2}.
Because the first two degrees of each divisor class~\eref{infclass2} are zero, in describing the superpotential contributions, we may restrict to the 3-dimensional subspace of $H_2(Y_4, \IC)$ spanned by $y_1$, $y_2$, and $y_3$, corresponding to the hyperplanes in the last three projective pieces of the ambient space, respectively. Then, the family~\eref{infclass2} contributes to the superpotential as
\begin{eqnarray}
W(y) &\propto& \sum_{a\in \mathbb Z} e^{2\pi i((-1+3a)y_1 + (1-2a)y_2 + (1-8a+10a^2) y_3 )} \nonumber \\  
&=& \sum_{a\in\mathbb Z} e^{2\pi i(a^2 \cdot 10 y_3 + a \cdot (3y_1 -2y_2 -8y_3) +(-y_1 + y_2 + y_3))} \nonumber\\
&=& e^{2 \pi i z} \sum_{a\in\mathbb Z} e^{2\pi i(\tau a^2 + w a) } \ ,  \label{partW}
\end{eqnarray}
where in the last step the following reparametrization has been made, 
\begin{eqnarray}
z&=&-y_1+y_2+y_3 \nonumber \ , \\
w&=&3y_1-2y_2-8y_3 \ , \\ 
\tau&=& 10y_3 \ \nonumber .
\end{eqnarray}
Now, to a Lie group $G$ of rank $r$ there is associated a theta function defined by
\beq
\Theta_{G}(\tau; \,\bold w=(w_1, \cdots, w_r)) \equiv \sum_{\bold m \in \Gamma_G} e^{2\pi i (\frac{\tau}{2} Q_G(\bold m) + \left<\bold m, \bold w\right> )} \ .
\eeq
$Q_G$ is the quadratic form associated with the Cartan matrix for the Lie algebra of $G$, and $\left<\cdot\,,\cdot\right>$ is the natural pairing. Then, the superpotential contribution, eq.~\eref{partW}, is proportional to $\Theta_{SU(2)}$ and hence, shows a modular behavior\footnote{It is important to observe that this modular behavior exists only because the pre-factors of each term in the series are identical in the case that the pulled-back divisors are sections to the elliptic fibration of $dP_9$. As pointed out in \cite{Donagi:1996yf} this is due to invariance under reparameterizing/shifting of elements of the Mordell-Weil group of the elliptic fibration. See \cite{Grimm:2015wda} for another look at the physical reparameterizations of Mordell-Weil elements.}. It should be noted that the one-parameter family of sections leading to the above superpotential is a subset of that which could be generated using the full rank 8 Mordell-Weil group, as discussed in \cite{Donagi:1996yf}. We leave to future work a systematic survey of the more general base surfaces of \cite{Morrison:2012np,Morrison:2012js,Martini:2014iza}, non-Higgsable clusters, and the intriguing question of how many such symmetries can appear in the non-perturbative superpotentials of dual heterotic/F-theory compactifications.


\section{Conclusions}\label{conc}
In this paper, we have studied instanton superpotential contributions from branes wrapping effective divisors of smooth CY manifolds, constructed as CICYs and  their generalizations,  gCICYs. In M-theory compactification on a smooth CY four-fold, instantons arise from M5-branes wrapping a divisor of the four-fold and may contribute to the superpotential of the effective $3D$, $\mathcal N=2$ theory. Via a network of string dualities, their consequences to the related $4D$, $\mathcal N=1$ theories are well established, where the relevant arenas are type IIB/F-theory (E3-brane instantons) or the heterotic theory (world sheet or spacetime instantons). 

Firstly, we have reviewed some known criteria for a divisor to potentially contribute to the superpotential. For a non-trivial contribution, the divisor necessarily has to be of arithmetic genus one (though this is not in general sufficient for a superpotential contribution), whereas a divisor being embedded rigidly is a sufficient condition (though not in general necessary). We have then provided explicit examples of CICY and gCICY geometries, together with the relevant brane-wrapping divisors that obey these criteria. Notably, the divisors are not inherited from the ambient space, but are nevertheless effective. 

In the network of dual theories in $4D$, fibered geometries determine much of the structure of the non-perturbative superpotential. In particular, for elliptically and/or K3 fibered CY four-folds, divisors that are pulled back from the base of the fibration play a special role. For the 2-dimensional base $B_2$ of a $K3$ fibration, the relevant $4D$ $\mathcal N=1$ duals are heterotic compactifications on an elliptically fibered CY three-fold with base $B_2$. Via the Leray spectral sequence, we have systematically studied the relevant cohomology structures and have thereby found that the sufficient rigidity criterion leads us to rational curves inside the base. In particular, amongst the ``minimal'' bases which generate the base surfaces of \cite{gross}, $dP_9$ turns out to be the only base that can give rise to an infinite family of rigid divisors on a non-trivial smooth CY four-fold. With such an infinite family of relevant divisors, the superpotential exhibits an intriguing modular behavior and we have indeed found a simple four-fold geometry with the $dP_9$ base, for which the resulting superpotential has an $SU(2)$ modular symmetry. Modular structures may also arise from CY geometries with non-minimal bases, and it would be interesting to explore them in a systematic manner, using the techniques that we have described here. 

It should be noted that in principle a systematic analysis of this same type can be applied to the 3-dimensional base $B_3$ of an elliptically fibered CY four-fold. The related $4D$ $\mathcal N=1$ theories in this case are IIB/F vacua. Furthermore, in the case that $B_3$ is a $\IP^1$ fibration over a surface $B_2$, the dual theories can also include Heterotic vacua with non-trivial superpotential terms generated via both world sheet instantons and spacetime instantons. In future work, we hope to add such effects to the systematic study of heterotic ``Standard Model" effective theories \cite{Anderson:2011ns,Anderson:2012yf,Anderson:2013xka,Anderson:2014hia} and their potentials \cite{Anderson:2011ty,Anderson:2013qca,Anderson:2009nt,Anderson:2011cza,Anderson:2014hia}. In the search for interesting infinite families of divisors, the Leray sequence remains a crucial tool. Unlike in the $B_2$ case, however, it is not straightforward to classify infinite families of relevant divisors on $B_3$ to uplift. In Appendix~\ref{lerayE}, similar criteria to those of $B_2$ case have been proposed for the lift of a base divisor to give rise to a desired divisor in $Y_4$. A detailed study of such examples (and the different modular symmetries they could give rise to) would be a fruitful area of future investigation.

In summary we have provided an improved preliminary toolkit for the study of divisors in CY four-folds and the associated non-perturbative superpotentials. We hope that in the future these tools can be extended to the context of realistic vacua in string phenomenology and for singular CY four-folds. In particular, the presence of such non-perturbative effects plays a critical role in moduli stabilization in the 4D, $\mathcal N=1$ theories. Explicit constructions such as those provided here can help to constrain the possible form of the moduli fixing potential, as well as shed light on the vacuum structure of the underling effective 3D $\mathcal N=2$ and 4D $\mathcal N=1$ theories arising from geometric engineering in string theory. 

\section*{Acknowledgements}
 LA and JG would like to thank the Aspen Center for Physics, which is supported by National Science Foundation grant PHY-1066293, for hospitality during the completion of this work. The work of LA (and XG in part) is supported by NSF grant PHY-1417337 and that of JG (and SJL in part) is supported by NSF grant PHY-1417316. FA would like to thank the City University of New York (Graduate Center) for hospitality and support, and the Columbia University for hospitality during part of this project. The work of FA is supported by the NSF Grant PHY-1452037. FA also acknowledges support from the Bahnson Fund at UNC Chapel Hill.
 
 \appendix
 
 \section{Leray Spectral Sequences for Elliptically and $K3$-fibered CY Four-folds}\label{leray}
In this section we consider CY four-folds that are elliptically and $K3$-fibered. We will consider each of these fibrations in turn and use divisors in the fibration bases to generate divisors $D \subset Y_4$ with arithmetic genus equal to one, satisfying \eref{strong_cond}.

For any fibered space, a Leray spectral sequence provides a simple tool to relate the cohomology of line bundles on the total space to some associated cohomology groups on the base. More specifically, let $\pi: Y \to B$ be a fibration with a generic fiber being $k$-dimensional and given by $\pi^{-1}(b)$ for a point in $b \subset B$.

Then we have a natural bi-grading such that for any bundle $V$ on $Y$,
\beq\label{spec_seq}
H^p(Y, V) = \sum_{p=l+m}E^{l,m}_{\infty} \ ,
\eeq
where 
\beq
E_{2}^{l,m}=H^l(B, R^{m}\pi_{*} (V)) \ ,
\eeq
and $R^{m}\pi_{*}(V)$ is the $m$-th direct image sheaf of the bundle $V$  (pushed forward under the fibration $\pi$).  The spectral sequence is iterated via the maps
\beq
d_r: E_{r}^{p,q} \rightarrow {E_{r}}^{p+r,q-r+1} \ ,
\eeq
where ${d_r}^2=0$ and 
\beq
E_{r+1}^{p,q}=\frac{ker(d_r: {E_{r}}^{p,q} \to {E_{r}}^{p+r,q-r+1})}{im(d_r: {E_{r}}^{p-r,q+r-1} \to {E_{r}}^{p,q})} \ ,
\eeq
where $E_{\infty}$ is defined as the limit to which this iterative sequence converges.

Note that on any open set $\mathcal{U}$ on $B$, the $m$-th direct image sheaf, $R^{m}\pi_{*} (V)$ can be locally represented by the pre-sheaf
\beq
\mathcal{U} \rightarrow H^m(\pi^{-1}(\mathcal{U}), V|_{\pi^{-1}(\mathcal{U})}) \ .
\eeq
It is clear then that if the fiber is $k$-dimensional, that $R^{m}\pi_{*} (V)$ is non-vanishingly only for $m=0, 1 \ldots k$.

To analyze the cohomology in any given fibration, a series of tools must be employed. The first of these will be useful to compute the cohomology of line bundles pulled back from the base. For any line bundle $\cO_B(D_B)$ on the base $B$, we can consider its pullback, $\pi^{*}(\cO_{B}(D_B))$. Then the push-forward functors $R^{m}\pi_{*}$ of this line bundle obey the so-called projection formula: For any bundles $V$ on $Y$ and $U$ on $B$,
\begin{equation}
R^{m}\pi_{*}(V\otimes \pi^{*}U)=R^{m}\pi_{*}(V)\otimes U \ . \label{projection}
\end{equation}
Another important tool is known as Grothendieck (or ``Relative") Duality \cite{AG,shafarevich}. For any sheaf ${\cal F}$ on $Y$, the push-forward functors obey the following relation:
\beq\label{grothendieck}
R^{k-i}\pi_{*}({\cal F}^{\vee}\otimes \omega_{Y|B})=(R^i\pi_{*}{\cal F})^{\vee}~~~,~~~i=0,1 \ldots k \ .
\eeq
where
\begin{equation}\label{dualsheaf}
\omega_{Y|B}= K_{Y}\otimes \pi^{*}({K_{B}}^{\vee}) \ ,
\end{equation}
and using the projection formula \eref{projection}, this reduces to
\beq
R^{k-i}\pi_{*}({\cal F}^{\vee}\otimes K_Y)\otimes K_{B}^{\vee}=(R^{i}\pi_{*} F)^{\vee} ~~~,~~~i=0,1 \ldots k \ .
\eeq

Using these tools, we will consider the following question. Let $\pi: Y \to B$ by a CY manifold. Suppose that a divisor $D_B \subset B$ has cohomology $h^0(B, \cO_B(D_B))=1$ and $h^i(B, \cO_B(D_B))=0$ $\forall i>0$. Under what conditions will this pull back to a rigidly embedded divisor on $Y$ with $h^i(Y, \pi^{*}(\cO_{B} (D_B)))=0$ $\forall i>0$ and $h^0(Y, \pi^{*}(\cO_{B} (D_B)))=1$? We will consider this in turn for $Y$ an elliptically, respectively $K3$, fibered four-fold.

\subsection{Elliptic fibrations, $\pi: Y_4 \to B_3$}\label{lerayE}
Let $\pi: Y_4 \to B_3$ be an elliptically (or genus one) fibered four-fold. Note that $R^{m}\pi_{*} (V)$ is non-vanishing only for $m=0,1$, because the fiber is 1-dimensional. Let $D \subset B_3$ have $h^0(B_3, \cO_{B_3}(D_{B_3}))=1$ and $h^i(B_3, \cO_{B_3}(D_{B_3}))=0$ $\forall i>0$ and define $L=\pi^*( \cO_{B_3}(D_{B_3}))$. Then it is clear that by the projection formula
\begin{align}
R^m\pi_{*} L= (R^m \pi_{*} \cO_{Y_4}) \otimes \cO_{B_3}(D_{B_3}) \ .
\end{align}
Furthermore, it will be useful to observe that using Grothendieck duality, \eref{grothendieck}, in this case:
\beq
R^0\pi_{*}(\cO_{Y_4})\otimes K_{B_3}^{\vee} = (R^1 \pi_{*}(\cO_{Y_4}))^{\vee} \ .
\eeq
Finally, for an elliptically fibered CY manifold it is straightforward to demonstrate that
\beq\label{trivial_pushforward}
R^0\pi_{*}(\cO_{Y_4})=\cO_{B_3}~~~~,~~~~~R^1\pi_{*}(\cO_{Y_4})=K_{B_3} \ .
\eeq

Now, with $L_{B_3} = \cO_{B_3}(D_{B_3})$, suppose that 
\beq
h^\bullet(B_3, L_{B_3} \otimes K_{B_3})=(0,0,k,k) \ ,
\eeq
for some integer $k \geq 0$, in which case the spectral sequence terminates at $E_2$. 
Then, the pullback bundle $L=\pi^*(\cO_{B_3}(D))$ has cohomology that is given by
\begin{align}
&H^0(Y_4,L)=H^0(B_3,R^0 \pi_{*}(L)) \ ,\\
&H^1(Y_4,L)=H^1(B_3,R^0 \pi_{*}(L))\oplus H^0(B_3, R^1 \pi_{*}(L)) \ ,\\
&H^2(Y_4,L)=H^2(B_3,R^0 \pi_{*}(L))\oplus H^1(B_3,R^1 \pi_{*}(L)) \ ,\\
&H^3(Y_4,L)=H^3(B_3,R^0 \pi_{*}(L))\oplus H^2(B_3,R^1 \pi_{*}(L)) \ ,\\
&H^4(Y_4,L)=H^3(B_3,R^1 \pi_{*}(L)) \ .
\end{align}
Using the projection formula, it is clear that $R^m\pi_{*}(L)=R^m\pi_{*} \cO_{Y_4} \otimes L_{B_3}$. Thus, by \eref{trivial_pushforward}, we have
\begin{align}\label{full_cohom}
&H^0(Y_4,L)=H^0(B_3,L_{B_3}) \ ,\\
&H^1(Y_4,L)=H^1(B_3,L_{B_3})\oplus H^0(B_3, L_{B_3} \otimes K_{B_3}) \ ,\\
&H^2(Y_4,L)=H^2(B_3,L_{B_3})\oplus H^1(B_3,L_{B_3} \otimes K_{B_3}) \ , \\
&H^3(Y_4,L)=H^3(B_3,L_{B_3})\oplus H^2(B_3,L_{B_3} \otimes K_{B_3}) \ , \\
&H^4(Y_4,L)=H^3(B_3,L_{B_3} \otimes K_{B_3}) \ .
\end{align}
Note that this result is manifestly consistent with Serre duality on $Y_4$ and on $B_3$, as expected. Thus, a line bundle/divisor, $L_{B_3}$ of the form described above will pull back to a rigidly embedded divisor with arithmetic genus equal to one, satisfying $h^\bullet(Y_4, L)=(1,0,0,k,k)$. Note that $k=0$ or $1$, as $k>1$ contradicts the Koszul sequence for $L$.

\subsection{$K3$ fibrations, $\rho: Y_4 \to B_2$}\label{lerayK3}
Let $\rho: Y_4 \to B_2$ be a $K3$ fibered four-fold. Here the fiber is two dimensional and $R^{m}\pi_{*} (V)$ is non-vanishing only for $m=0,1,2$.
As in the previous case, consider $D_{B_2} \subset B_2$ with $h^0(B_2, \cO_{B_2}(D_{B_2}))=1$ and $h^i(B_2, \cO_{B_2}(D_{B_2}))=0$ $\forall i>0$ and define $L=\rho^*(\cO_{B_2}(D_{B_2}))$. As in the previous section, by the projection formula, \eref{projection}, the cohomology of $L$ on $Y_4$ is fully specified by the higher derived push-forward functors of the trivial line bundle $\cO_{Y_4}$:
\beq
R^{m}{\rho}_{*}(\cO_{Y_4}) ~~~~\text{for}~~~m=0,1,2 \ .
\eeq
For a $K3$-fibered CY four-fold, it can be shown that
\beq
R^{0}{\rho}_{*}(\cO_{Y_4}) =\cO_{B_2}~~~~,~~~~R^{1}{\rho}_{*}(\cO_{Y_4})=0,~~~~~,~~~~~R^{2}{\rho}_{*}(\cO_{Y_4})=K_{B_2} \ .
\eeq
With $L_{B_2} = \cO_{B_2}(D_{B_2})$, presuming again that 
\beq
h^\bullet(B_2, L_{B_2} \otimes K_{B_2})=(0,k,k) \ ,
\eeq
for some integer $k\geq 0$, we see that the spectral sequence terminates at $E_2$ and that the cohomology forms a pattern very similar to that given in the previous section,
\begin{align}\label{full_cohom_K3}
&H^0(Y_4,L)=H^0(B_2,L_{B_2}) \ ,\\
&H^1(Y_4,L)=H^1(B_2,L_{B_2}) \ ,\\
&H^2(Y_4,L)=H^2(B_2,L_{B_2})\oplus H^0(B_2,L_{B_2} \otimes K_{B_2}) \ , \\
&H^3(Y_4,L)=H^1(B_2,L_{B_2} \otimes K_{B_2}) \ , \\
&H^4(Y_4,L)=H^2(B_2,L_{B_2} \otimes K_{B_2}) \ .
\end{align} 
We then find that $h^\bullet(Y_4, L)=(1,0,0,k,k)$ where $k$ has to be either $0$ or $1$ again for a consistency with the Koszul sequence for $L$.

\end{document}